\documentclass[]{raa}

\usepackage{graphicx,times}             
\usepackage{natbib}
\usepackage{amssymb,amsmath}

\bibpunct{(}{)}{;}{a}{}{,}

\usepackage[a4paper=true,dvipdfm=true,pagebackref=true]{hyperref}
\hypersetup{colorlinks = true, linkcolor = green, anchorcolor = red, citecolor = blue, filecolor = red, pagecolor = red, urlcolor = red}

\begin{document}

   \title{Long-term multi-wavelength variations of Fermi blazar 3C 279
$^*$
\footnotetext{$*$ Supported by the National Natural Science Foundation of China.}
}

   \volnopage{Vol.0 (20xx) No.0, 000--000}      
   \setcounter{page}{1}          

   \author{Bing-Kai Zhang
      \inst{1,2}
   \and Min Jin
      \inst{1}
   \and Xiao-Yun Zhao
      \inst{1}
   \and Li Zhang
      \inst{3}
   \and Ben-Zhong Dai
      \inst{3}
   }

   \institute{Department of Physics, Fuyang Normal University, Fuyang 236037, China; {\it zhangbk\_ynu@163.com}\\
        \and
             Key Laboratory of Functional Materials and Devices for Informatics of Anhui Higher Education Institutes, Fuyang Normal University, Fuyang 236037, China\\
        \and
             Department of Astronomy, Yunnan University, Kunming 650091, China\\
\vs\no
   {\small Received~~20xx month day; accepted~~20xx~~month day}}

\abstract{ The long-term optical, X-ray and $\gamma$-ray data of blazar 3C 279 have been compiled from $Swift$-XRT, $RXTE$ PCA, $Fermi$-LAT, SMARTS and literature. The source exhibits strong variability on long time scales. Since 1980s to now, the optical $R$ band light curve spans above 32 yr, and a possible 5.6-yr-long quasi-periodic variation component has been found in it. The optical spectral behavior has been investigated. In the optical band, the mean spectral index is -1.71. The source exhibits an obvious special spectral behavior. In the low state, the source shows a clear bluer-when-brighter behavior in a sense that the optical spectrum turns harder (flatter) when the brightness increases.  While in the high state, the optical spectrum is stable, that means the source spectral index does not vary with the brightness. The correlation analysis has been performed among optical, X-ray and $\gamma$-ray energy bands.
The result indicates that the variations of $\gamma$-ray and X-ray bands are well correlated without time delay on the time scale of days, and their variations exhibit weak correlations with those of optical band. The variations, especial outbursts, are simultaneous, but the magnitude of variations is disproportionate. The detailed analysis reveals that the main outbursts exhibit strong correlations in different $\gamma$-ray, X-ray and optical bands.
}
   \authorrunning{B.-K. Zhang et al. }            
   \titlerunning{Multi-wavelength variations of 3C 279}  

   \maketitle

%
%
\section{Introduction}           
\label{sect:intro}

Blazars, comprising BL Lacs and FSRQs, are the most extreme subclass of active galactic nuclei (AGNs) with a relativistic jet pointing along to our line of sight \citep{urry1995}.  They radiate out in all electromagnetic bands from radio to $\gamma$-rays, and exhibit non-thermal emission. In general, there are two typical broad peaks in their spectral energy distributions (SEDs). One is low energy peak which lies between infrared and optical bands, even extending to X-ray band, and another is high energy peak which locates from MeV to TeV $\gamma$-ray. The low energy peak can be interpreted by synchrotron radiation from relativistic energy electrons in the jet, while the mechanism of the second peak is an open question. At present, two typical models, leptonic model and hadronic scenario, are used to describe the high energy peak.
Blazars exhibit violent variability from radio band to very high energy $\gamma$-ray band on diverse time scales which range from years to days, event to several minutes. So, the variability study is an important tool to provide valuable information about their nature.

The 3C 279,  known as a flat spectrum radio quasar (FSRQ) with $z$ = 0.538, is the first quasar which was discovered to exhibit apparent super-luminal motion \citep{whitney71,unwin89}. The apparent speeds can range up to $\beta_{app} \sim 17$ \citep{jorstad04}.
3C 279 is one of the first blazars which have been detected to emit $\gamma$-rays \citep{hartman92}. And it is the first FSRQ discovered to emit very high energy $\gamma$-rays \citep{albert08}.
Since 2008, 3C 279 has been observed continuously at high energy $\gamma$-rays (HE, E $>$ 100 MeV) by the \emph{Fermi} satellite.
 The shortest doubling time scale of 1.19 hr has been reported in $\gamma$-ray variability during  2014 March-April \citep{paliya15}.

The variable behavior of 3C 279 has been widely studied in the optical band. It shows very large and rapid variations in brightness. A very large and violent optical
outburst from $B$ = 18 to 11.27 mag in $\sim$1.5 yr was reported by \cite{eachus75}. After 1951, the source became less active. And then, since 1987, the source exhibited more active variation. A rapid outburst of 2 mag within 24 hr was reported \citep{webb90}, and the most rapid optical $V$ band variation by 1.17 mag in 40 min was observed on 1996 May 22 by \cite{xie99}. During 2001-2002, the source exhibited optical $R$ band variations up to 0.5 mag in 24 hr and variations up to 0.13 mag in 3 hr \citep{kartaltepe07}.
A strong flux decline ($\Delta R > 1.1$ mag in 13 d) was reported according to the WEBT campaign during 2005-2006 \citep{bottcher07}.
With infrared $K$-band data, a strong period of 7.1 $\pm$ 0.44 yr was reported by \cite{fan99}. In the optical $R$ light curve, there seemed to be a 256-d quasi-periodicity signal \citep{sandrinelli16}.

The source 3C 279 has been intensively studied through several simultaneous multi-wavelength campaigns \citep{hartman96,wehrle98,bottcher07,shah19,larionov20,prince20},
and the correlation between the variations of different energy bands has been investigated.
For example,
A clear radio-optical correlation with 0-35 d time lag was found by \cite{tornikoski94} and \cite{zhang17}.
Between optical and near-IR bands, the correlation was seen historically in 3C 279 \citep{hartman96,wehrle98}.
Between infrared and gamma-rays, \cite{bonning12} found weak correlations.
\cite{hartman01} investigated cross-correlations among optical, X-ray, and $\gamma$-ray bands, and found no consistent tendencies although a significant optical/$\gamma$-ray correlation was found during 1999, with an $\sim$2.5-d $\gamma$-ray lag.
\cite{chatterjee08} presented that variations were significantly correlated between radio, optical and X-ray bands.
Using the first two-year $Fermi$ data, \cite{hayashida12} found $\gamma$-ray emission preceded optical one by about 10 d. However, they found a lack of correlation between variations in X-ray and $\gamma$-ray bands during 2008-2010. They also detected a X-ray flare with no obvious counterpart in other energy bands.
On the contrary, the different trends in 2011 May and June were reported by \cite{aleksi14} that X-rays were correlated with $\gamma$-rays whereas the optical band seemed to be no significant correlation with $\gamma$-rays. With decade-long (2008-2018) data, \cite{larionov20} also concluded that the X-rays were well correlated with $\gamma$-rays with no lag being greater than 3 hr, and the $\gamma$-ray flux presented a complex relationship with optical flux, changing with the state of the source activity. During the flare of 2017-2018, 3C 279 showed an obvious correlation between $\gamma$-rays and X-rays with no delay \citep{prince20}.

With the accumulation of observed data of 3C 279, it is very meaningful to research the long-term multi-band variability behavior of this source. The outline of the paper is arrayed as follows. The second section describes the light curves of optical, X-ray and $\gamma$-ray bands of 3C 279. Then in the third section, we search for the quasi-periodic variable signals in the optical light curve. After this, we analyze the optical spectral behavior in the fourth section, and investigate cross correlations between three different energy bands in the fifth section. Finally we give a discussion and summarize our findings in the sixth section.

\section{Light curves}

\begin{figure}
   \centering
   \includegraphics[width=\textwidth, angle=0]{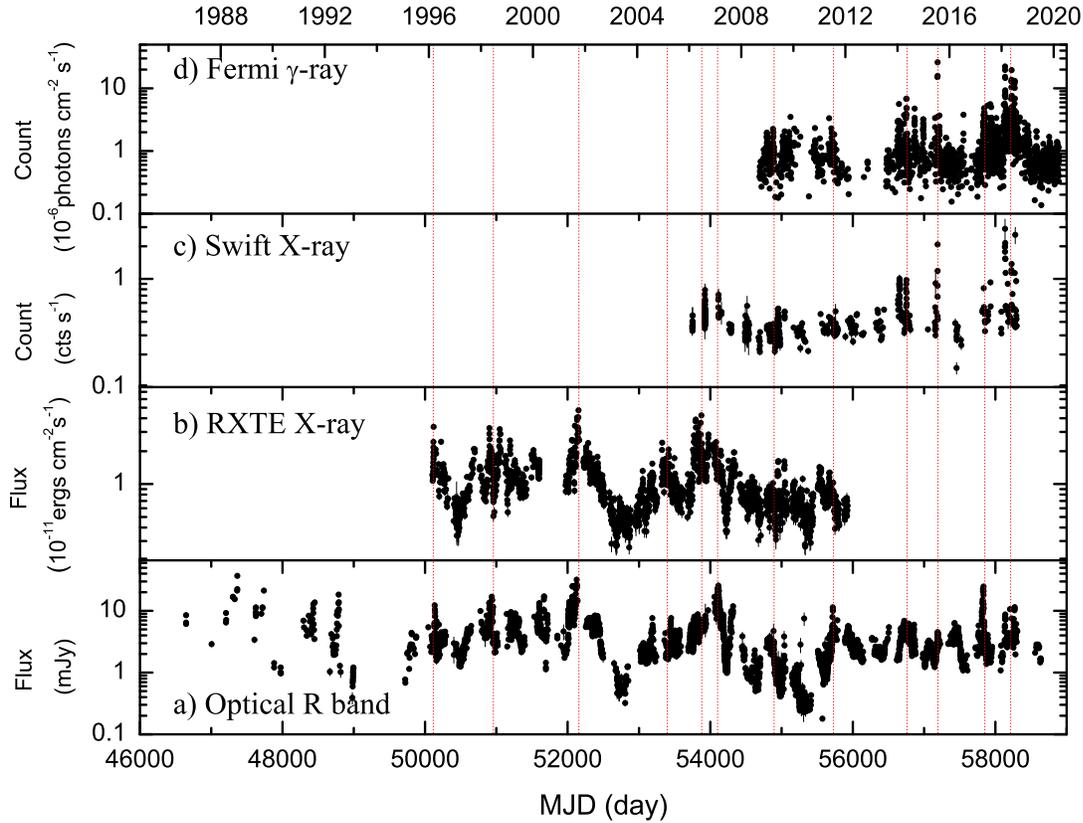}
   \caption{The long-term different energy band light curves of blazar 3C 279. From panel (a) to (d), they represent the light curves of optical ($R$ band), $RXTE$ X-ray (2-10 KeV), $Swift$ X-ray (0.3-10 KeV) and $Fermi$ $\gamma$-ray (0.1-300 GeV) energy bands, respectively. The vertical red dot lines have been drawn near the main flares to guide the eyes. Calendar dates are along the top and Modified Julian Dates are shown at the bottom.
   \label{lightcurve}}
\end{figure}

\begin{figure}
\centering
   \includegraphics[width=10cm, angle=0]{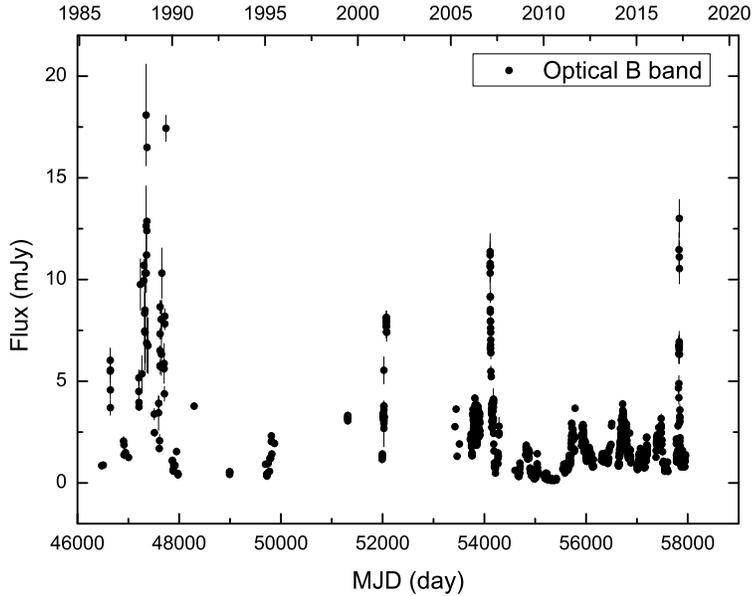}
   \caption{The long-term optical $B$ band light curve of blazar 3C 279. Calendar dates are along the top and Modified Julian Dates are shown at the bottom.
   \label{lightcurveb}}
\end{figure}

\subsection{Optical Band}
We have compiled the observed data in the optical $R$ band from different literature
\citep{chatterjee08,dai01,ghosh00,grandi96,hartman96,hayashida12,kartaltepe07,katajainen00,kidger92,kiehlmann16,
larionov08,mead90,Nilsson18,sandrinelli16,shrader94,takalo92,villata97,webb90,xie99,xie01,xie02}
and some groups, such as  the Small and Moderate Aperture Research Telescope System \citep{bonning12}, the Steward Observatory monitoring project \citep{smith09} and the 0.76-m Katzman Automatic Imaging Telescope \citep{li03}. Totally, there are 7299 data points in the optical $R$ band light curve.  The length of the data series is over a span of 32 yr, from MJD 46646 to 58306. The light curve has been presented in panel (a) of Figure~\ref{lightcurve}.
The figure shows that the source exhibits some large outbursts near MJD 52126, 54113 and 57834. There is also a huge outburst with $F_{R}$ = 121.5 mJy ($R$ = 11.01 mag) on 47306. Due to the scale, it is not showed in the figure. Additionally, there are still some slightly smaller flares. It seems that the source bursts more frequently before MJD 52000 than after. During the past 3 decades, the source varies violently with $\Delta R$ = 7.08 mag. Meanwhile, we have collected the data of optical $B$ band, which consists of 1083 points (see Fig.~\ref{lightcurveb}). \cite{webb90} observed a big outburst in 1988 with $B$ = 12.13 mag. Even though, it was still 0.83 mag fainter than the one during 1936-1937 ($B$ = 11.3 mag).

\subsection{$RXTE$ X-ray}
The Rossi X-ray Timing Explorer ($RXTE$) sustained long-term X-ray observations of sources during its mission from 1995 December to 2012 January, and it provided 2-10 keV light curves from the Proportional Counter Array (PCA) data with one point per observation \citep{rivers13}. 3C 279 was continually monitored with 2-3 points per week from MJD 50104 (1996 January 22) to 55925 (2011 December 30), and the total good exposure time of PCA is 2222 ks. The light curve includes 1987 observations and data points with a time coverage about 15 yr $\footnote{https://cass.ucsd.edu/$\sim$rxteagn/}$. It is plotted in panel (b) of Figure ~\ref{lightcurve}. One can see that it exhibits 4 main outbursts in the light curve as well as some low amplitude flares.

\subsection{$Swift$ X-ray}
3C 279 was observed by X-ray Telescope (XRT) equipped on the satellite of $Swift$ from MJD 53748 to 58299 with 6.89$\times10^{5}$ seconds of exposure time, which spans about 12.5 yr. The data set is comprised of 471 points$\footnote{https://www.swift.psu.edu/monitoring}$. The analysis methods used to produce X-ray light curves were described by \citep{stroh13}. The light curve has been shown in panel (c) of Figure~\ref{lightcurve}. One can see that the 3C 279 exhibits the largest outburst on MJD 58135 (2018 January 17) since it was observed in 2006 January. Then, 143 d after this outburst, a followed one appeared on MJD 58278 (2018 June 9). Another outburst can be seen on MJD 57189 (2015 June 6). In addition, there are some flares in the light curve accompanied by some gaps.

\subsection{$Fermi$ $\gamma$-ray}
$Fermi$-LAT is a $\gamma$-ray telescope which is sensitive to the energy of 20 MeV - 300 GeV \citep{atwood09}.
$Fermi$ covers the entire sky every 90 minutes.
3C 279 has been observed from MJD 54691 (2008 August 13) to 58893 (2020 February 14) with a time span of about 11.5 yr. Its daily flux has been provided by $Fermi$-LAT$\footnote{https://fermi.gsfc.nasa.gov/ssc/data/access/lat/msl\_lc/}$. The light curve is comprised of 130 data points and is drawn in panel (d) of Figure~\ref{lightcurve}. From the light curve, one can see an obvious outburst on MJD 57189 (2015 June 6). During the outburst, the flux increased by 12.5 times in 5 d, then decreased by 97.3\% in 5 d. There are two other large outbursts on MJD 58136 (2018 January 18) and 58228 (2018 April 20), respectively. This means that 3C 279 is a high variable source in $\gamma$-ray band.

\begin{figure}
  \centering
  \includegraphics[width=10cm, angle=0]{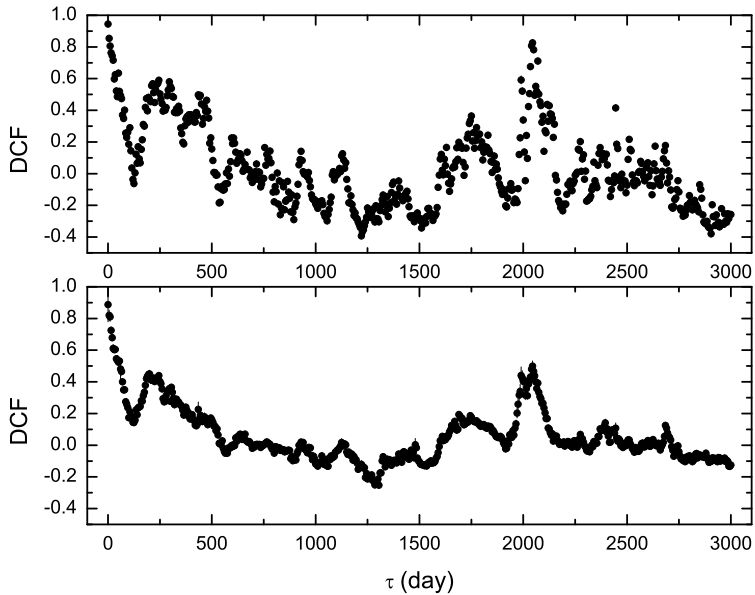}
  \caption{The auto-DCF of the optical $R$ band light curve of 3C 279. The upper is the auto-DCF for the full light curve, and the lower is the auto-DCF for one-day averaged light curve.
  \label{dcfrr}}
\end{figure}

\section{Periodicity}

\subsection{DCF Method}
The DCF (discrete correlation function) method was presented by \cite{edelson88} to search for correlations and the possible time delays between two different data sets. The method can also be used to search for possible periodic component in a single light curve. This method takes full advantage of all available data points, and no interpolation is required in the light curve. So it is peculiarly propitious for unevenly sampled variability data.
Given a time delay $\tau$, all data pairs with time delay of $\Delta t_{ij}$ $\in [\tau-\Delta \tau/2, \tau+\Delta\tau/2)$ are employed to compute the DCF$_{\tau}$
\begin{equation}
DCF(\tau)=\frac{<(a_{i}-\overline{a})(b_{j}-\overline{b})>}{\sigma_{a}\sigma_{b}}
\end{equation}
for two light light curves $a$ and $b$, where $a_i$ and $b_j$ are the fluxes of data points,
$\overline{a}$ and $\overline{b}$ are the flux means, and $\sigma_a$ and $\sigma_b$ are the standard deviations, respectively \citep{white94,fuhrmann14}.

For each time delay, $\tau$, there is a DCF corresponding to it.
The DCF peak means two light curves are correlated with each other, and exist a delay of $\tau$.
For a single light curve, the evident peak of auto-DCF indicates the existence of periodic signals.
In general, Monte Carlo simulation method is used to evaluate the time delay and its error \citep{peterson98,raiteri03}.

\subsection{Periodicity Analysis}
The long-term optical $R$ band light curve of 3C 279 is searched by the DCF method to find possible periodic or quasi-periodic signals.
The DCF result is plotted in Figure~\ref{dcfrr}.  Besides a peak at zero, there exists an obvious peak at 2045 d with DCF $=$ 0.83 $\pm$ 0.01. 5000 simulations have been performed, and the peak position as well as its uncertainties are derived to be 2044 $\pm$ 2.1 d.
To suppress the spurious signals, the light curve are daily averaged. The corresponding DCF is also plotted in  Figure~\ref{dcfrr}. The peak is still at $\tau=$ 2045 d, but much lower, which equals 0.50 $\pm$ 0.03. Light curve simulations show that its significance is 4.1$\sigma$.
This means the light curve has a very weak 5.6-yr quasi-periodic variation component.

From the light curve, one can see several obvious outbursts. The result of $\tau$ = 5.6 yr may be dominated by some outburst pairs, which have about 5.6-yr-long intervals, such as, the outbursts on MJD 57834 and 55719 with an interval of 5.79 yr, the outbursts on MJD 55719 and 54113 with an interval of 4.4 yr, the outbursts on MJD 54113 and 52126 with an interval of 5.44 yr, the pairs on MJD 52126 and 50136 with an interval of 5.45 yr, and the pairs on MJD 50936 and 48789 with an interval of 5.88 yr. It should be noted that there is a huge outburst on MJD 47306 without a counterpart one.

\begin{figure}
  \centering
  \includegraphics[width=10cm, angle=0]{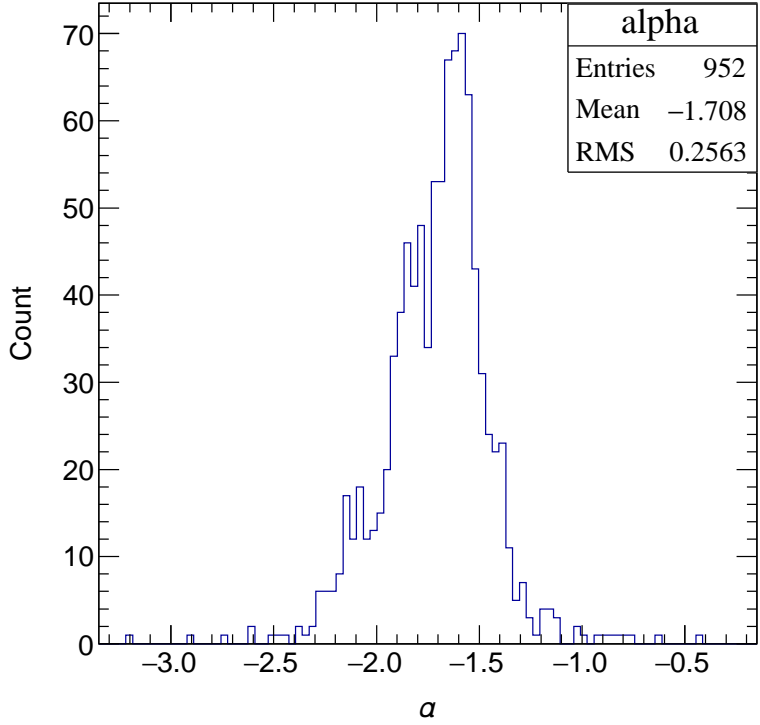}
  \caption{The distribution of optical spectral indices $\alpha$s.
  \label{alphadis}}
\end{figure}

\section{Spectral behavior investigation}
The spectrum generally follows a power law for blazars, that is,
\begin{equation}
  F\propto\nu^{\alpha},
\end{equation}
where $F$ represents the flux, $\nu$ represents the frequency, and $\alpha$ represents the spectral index.

Then the spectral index can be computed as
\begin{equation}
  \alpha=\frac{logF_{2}-logF_{1}}{log\nu_{2}-log\nu_{1}}
  \label{equ-alpha}
\end{equation}
 where subscripts 1 and 2 denote two different observation bands.

The optical variations are always accompanied by spectral variations. To investigate the spectral behavior, we have computed the optical spectral index between optical $B$ and $R$ bands.  Quasi-simultaneous $B-R$ pairs of observations with intervals less than 30 min are chosen for the analysis. In addition, a correction for interstellar extinction are introduced. For 3C 279, the color excess due to interstellar extinction is $E_{B-V}$ = $0.03^{m}$ \citep{mead90}, and then the interstellar extinction $A_{V}$ in the $V$ band is $0.093^{m}$. Using the normal interstellar extinction curve \citep{schlegel98}, we obtain $B$ and $R$ band extinctions. The values of $A_{B}$ and $A_{R}$ are $0.123^{m}$ and $0.078^{m}$, respectively.  After correction, the $\alpha$s are calculated with Equation~\ref{equ-alpha}. The distribution of $\alpha$s is given in Figure~\ref{alphadis}, which shows the $\alpha$s have a mean value of $-1.71\pm0.26$.

\begin{figure}
  \centering
  \includegraphics[width=10cm, angle=0]{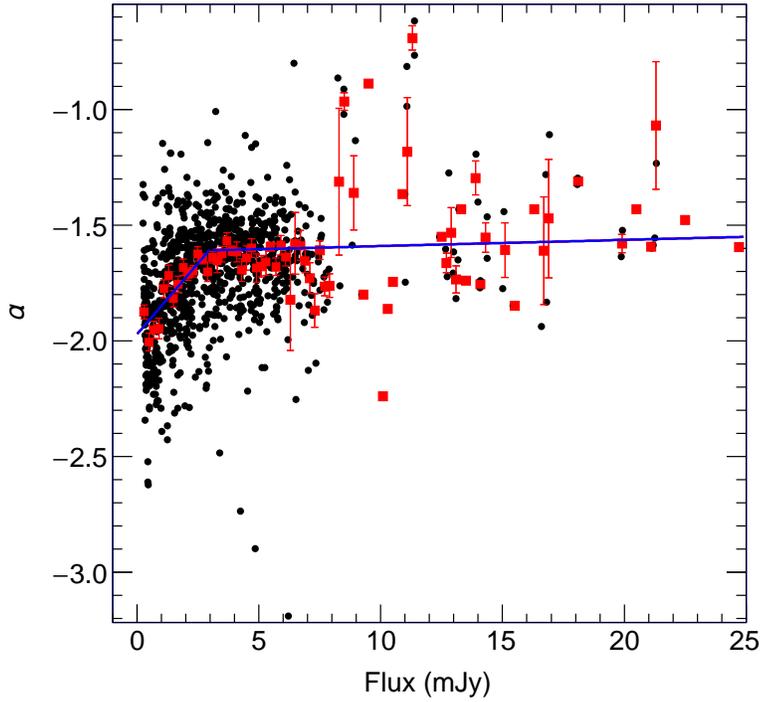}
  \caption{Optical spectral index versus $R$ band flux. The red squares with error bars represent the average values of $\alpha$s. The two blue lines indicate the linear fitting results.
  \label{alpha}}
\end{figure}

The change of spectral index with flux is presented by black dots in Figure~\ref{alpha}.
One can see that these points are very scattered. However, they have a gradual rising trend from the overall outline, which means the spectrum becomes flatter as the flux increases.
However, if we see the details, there seem to be a fine structure. To see clearly, we calculate the average values of $\alpha$s at different fluxes, and superimpose them on the Figure~\ref{alpha} with red squares and error bars. An obvious variation tendency can be clearly seen. The spectrum index increases from $-2.0$ to $-1.6$ with the brightness increases, and then hardly increases any more when flux reaches a certain value around 3 mJy (15.03 mag).

The data are divided into two segments and fitted with two different linear functions. The fitted results are also superimposed on Figure~\ref{alpha} with two blue lines. When fluxes of $R$ band, $F_{R}$, are below 3 mJy,
 \begin{equation}
 \alpha=(0.123\pm0.011)F_{R} -(1.970\pm0.020),
 \end{equation}
and when above 3 mJy,
 \begin{equation}
 \alpha=(0.003\pm0.001)F_{R} - (1.617\pm0.009).
 \end{equation}
The fitting results show that in the low state of the source, $\alpha$ increases obviously with the increase of brightness, and then stays roughly stable in the high state.

\begin{figure}
  \centering
  \includegraphics[width=10cm, angle=0]{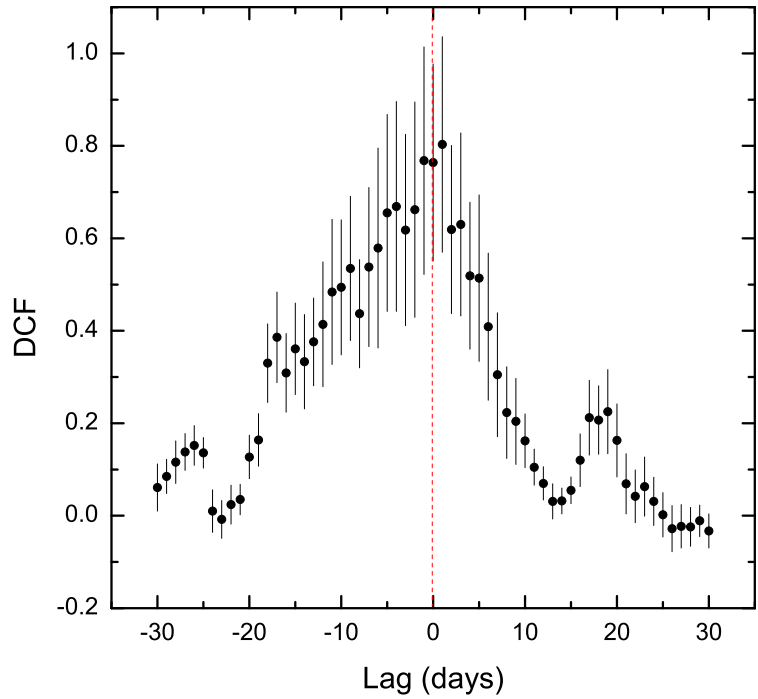}
  \caption{The DCF between $Fermi$ $\gamma$-ray and $Swift$ X-ray bands of 3C 279. The red vertical dashed line is used to highlight the position of zero.
  \label{dcf-fermi-swift}}
\end{figure}

\begin{figure}
  \centering
  \includegraphics[width=10cm, angle=0]{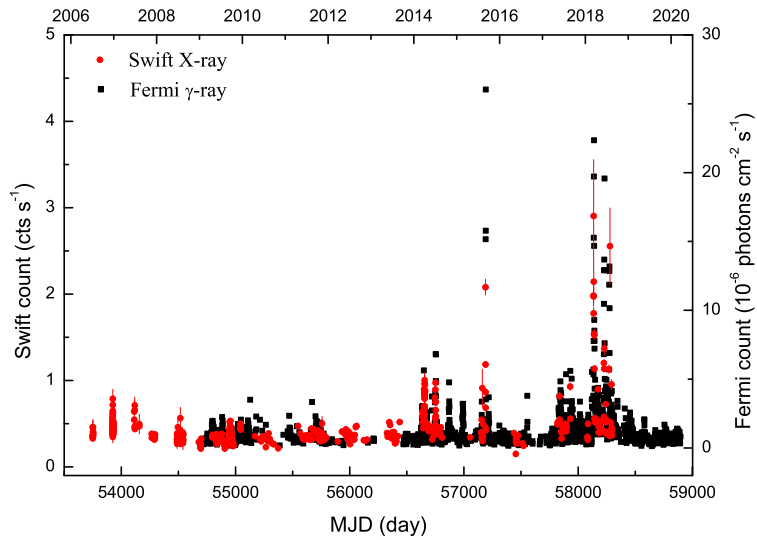}
  \caption{The light curves of $Fermi$ $\gamma$-ray (black squares) and $Swift$ X-ray (red dots) bands.
  \label{lc-fermi-swift}}
\end{figure}

\section{Correlation analysis}

\subsection{Correlation between $\gamma$-ray and X-ray Bands}
The light curves of $\gamma$-ray and X-ray (observed by $Swift$-XRT) have about 9.9 yr of overlap from MJD 54691 to 58299 (see panels (d) and (c) in Fig.~\ref{lightcurve}). We apply the DCF method to analyze the relation between the two light curves with $\Delta\tau$ = 1 d.
An obvious peak (DCF = 0.80 $\pm$ 0.23) can be seen at $\tau$ = 1 d (Fig.~\ref{dcf-fermi-swift}). We have performed 5000 Monte Carlo simulations, and derived $\tau_{center}$ = 0.0$^{+1.0}_{-1.8}$ d and DCF = 0.85$^{+0.05}_{-0.06}$,  which implies that $\gamma$-ray variations are well correlated with X-ray ones without time lags. To see clearly the variation trends, we have plotted these two light curves on the same panel in Figure~\ref{lc-fermi-swift}. The light curves have the similar structures and trends, and they vary synchronously (in step), especially during the main outbursts during 2015 June, 2018 January and 2018 April.

\begin{figure}
  \centering
  \includegraphics[width=10cm, angle=0]{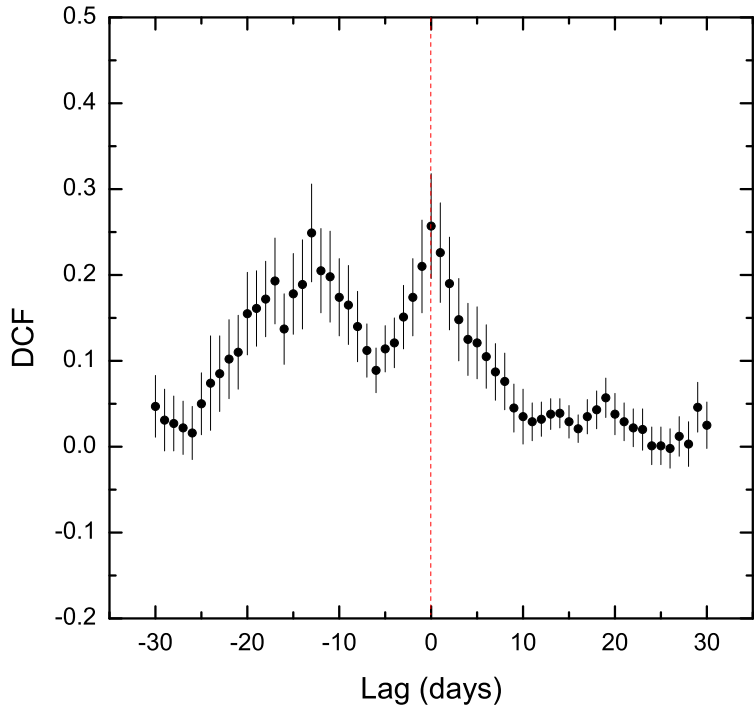}
  \caption{The DCF between $Fermi$ $\gamma$-ray and optical $R$ bands. The red vertical dashed line is used to highlight the position of zero.
  \label{dcf-fermi-optical}}
\end{figure}

\begin{figure}
  \centering
  \includegraphics[width=10cm, angle=0]{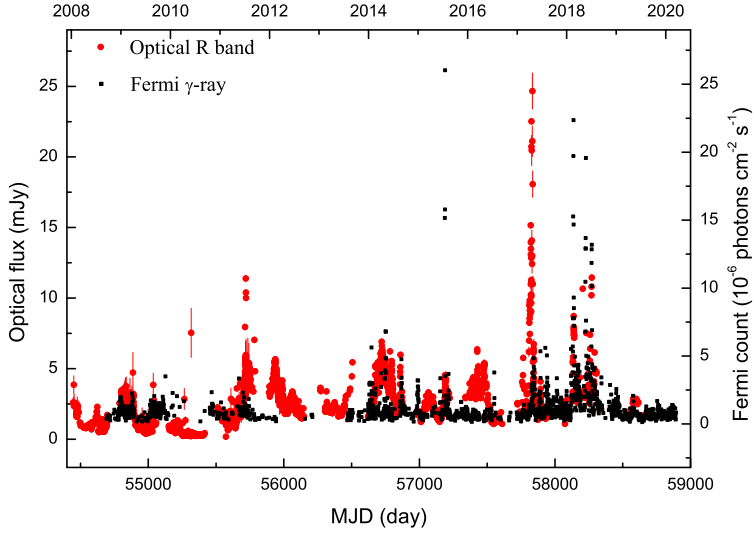}
  \caption{The light curves of $Fermi$ $\gamma$-ray (black squares) and optical $R$ (red dots) bands.
  \label{lc-fermi-optical}}
\end{figure}

\subsection{Correlation between $\gamma$-ray and Optical Bands}
Light curves of $\gamma$-ray and optical $R$ bands have a 10.8-yr-long overlap from MJD 54691 to 58637 (see panels (d) and (a) in Fig.~\ref{lightcurve}).  The correlation between them is investigated by DCF method. Figure~\ref{dcf-fermi-optical} presents the result, which has a peak of 0.26 $\pm$ 0.06 at zero as well as another peak of 0.25 $\pm$ 0.06 at $-13$ d. Using 5000 simulations, we have derived $\tau_{center}$ to be $-0.5^{+1.0}_{-12.5}$ d, and DCF to be $0.31^{+0.05}_{-0.05}$. This suggests that variations of $\gamma$-ray band are weak correlated with those of optical bands. From panels (a) and (d) in Figure~\ref{lightcurve}, one can see that almost each flare in $\gamma$-ray band has a counterpart in optical band. Two light curves are superimposed on each other, and are drawn in Figure~\ref{lc-fermi-optical} to exhibit the details of variations. One can see that the corresponding scintillation of two light curves occur almost at the same time (simultaneously), but the amplitude of variations are not very proportional. For example, in the light curve of optical band, the burst on MJD 57834 is greater than that on MJD 58271.  In the $\gamma$-ray region, however, the opposite is true.

\subsection{Correlation between X-ray and Optical Bands}

\begin{figure}
  \centering
  \includegraphics[width=10cm, angle=0]{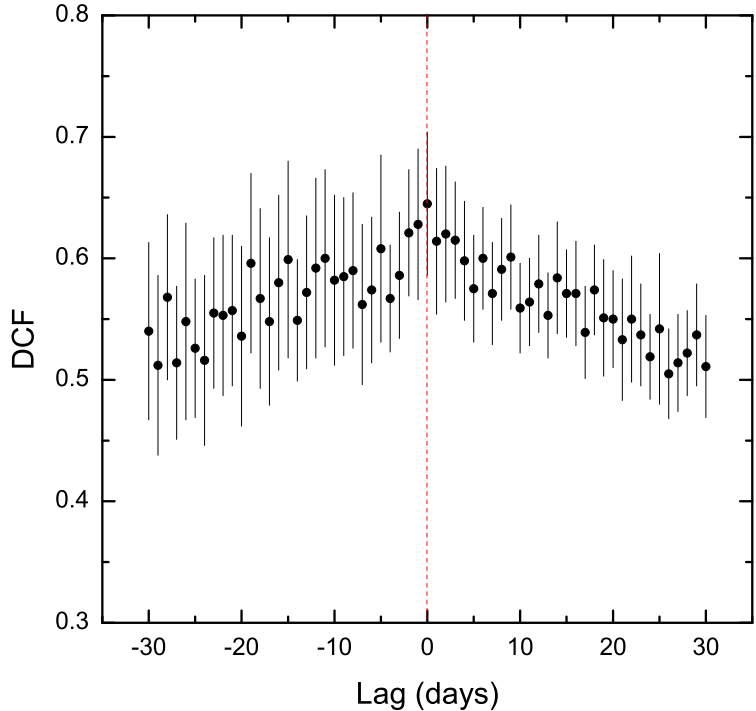}
  \caption{The DCF between $RXTE$ X-ray and optical $R$ bands. The red vertical dashed line is drawn to guide the eyes.
  \label{dcf-rxt-opt}}
\end{figure}

\begin{figure}
  \centering
  \includegraphics[width=10cm, angle=0]{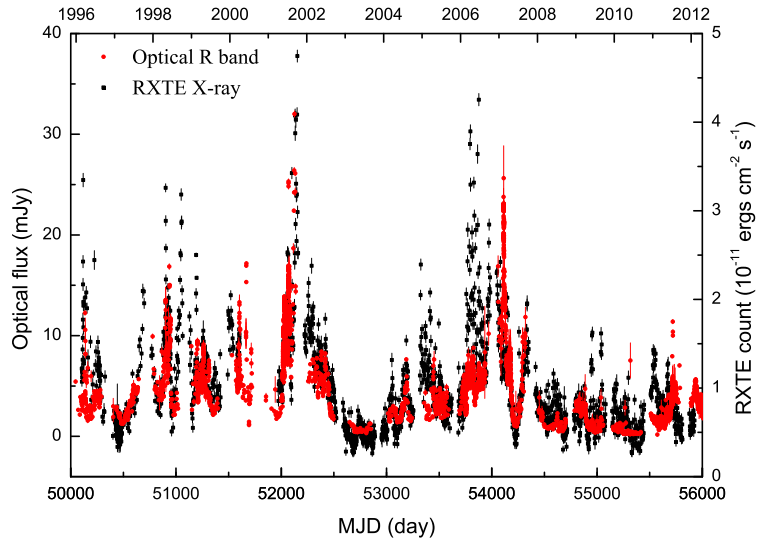}
  \caption{The light curves of $RXTE$ X-ray (black squares) and optical (red dots) bands. Only partial optical light curve has been plotted.
  \label{lc-optical-rxte}}
\end{figure}

The results of DCF, between X-ray (observed by $RXTE$-PCA) and optical $R$ light curves, are presented in Figure~\ref{dcf-rxt-opt}. There is an obvious peak (DCF = 0.65 $\pm$ 0.06) at zero. To estimate the error of time delay, we have performed 5000 simulations, and obtained $\tau_{center}$ = $0.0^{+4.8}_{-4.4}$ d with DCF = $0.67^{+0.03}_{-0.03}$. This means that the light curve of X-ray band correlates well with that of the optical band with zero lag. In light curves plotted in Figure~\ref{lightcurve}, we have drawn some vertical red dot lines near the peaks to guide the eyes. The two light curves exhibit flares simultaneously. Four main peaks (near MJD 54113, 52126, 50936 and 50136) in the light curve of optical band all have counterparts in $RXTE$ X-ray light curve. However, the X-ray flare in early 2006 seems to be no optical counterpart. To see details of variations, we have superimposed the light curve of optical band on that of $RXTE$ X-ray band, and shown them in the same panel (Fig.~\ref{lc-optical-rxte}). It is quite apparent that the variation trends are very similar to each other.

\begin{figure}
  \centering
  \includegraphics[width=10cm, angle=0]{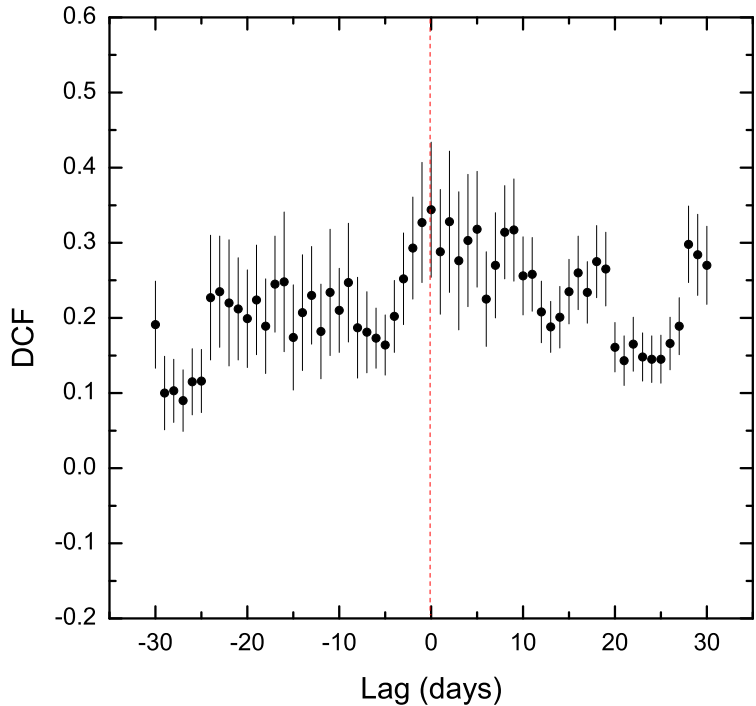}
  \caption{The DCF between $Swift$ X-ray and optical $R$ bands. The vertical red dashed line is for guiding the eyes.
  \label{dcf-swi-opt}}
\end{figure}

\begin{figure}
  \centering
  \includegraphics[width=10cm, angle=0]{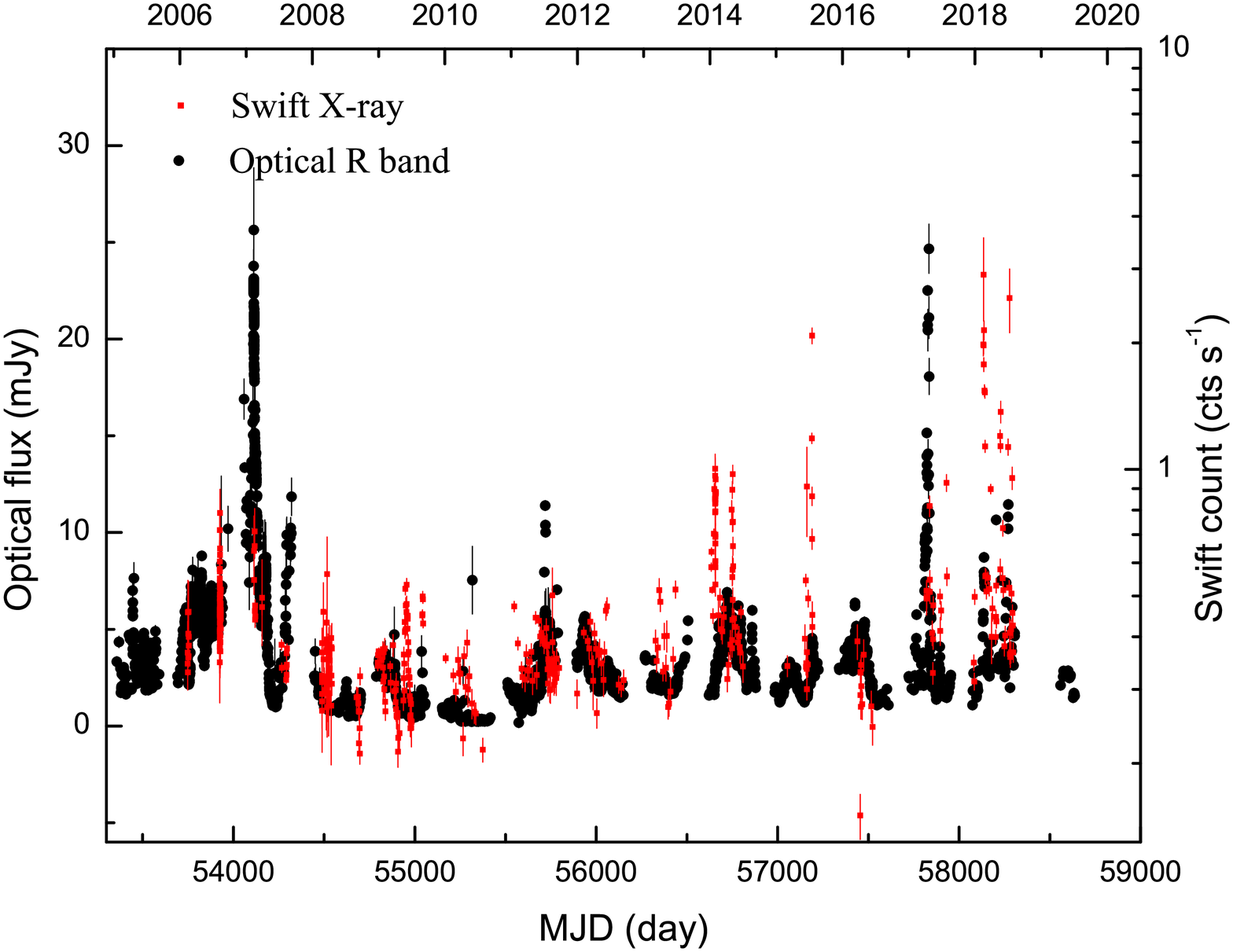}
  \caption{The light curves of $Swift$ X-ray (red squares) and optical (black dots) bands.
  \label{lc-swift-optical}}
\end{figure}

There is a 12.5-yr overlap between $Swift$ X-ray and optical light curves from MJD 53748 to 58299. We have also investigated the correlation between two light curves. The DCF result is given in Figure~\ref{dcf-swi-opt}. There is a peak with DCF = 0.34 $\pm$ 0.09 at $\tau$ $=$ 0 d. According to 5000 simulations, the mean value of time delay is derived to be 1.9$^{+7.0}_{-3.5}$ d, and DCF = 0.43$^{+0.06}_{-0.05}$.
The two curves are superimposed on each other, which have been shown in Figure~\ref{lc-swift-optical}. The source bursts almost simultaneously, and the light curves have similar trends. However, the magnitude of the main flares are disproportionate. For example, around MJD 57837, the X-ray burst looks lower than its counterpart in the optical band, whereas around MJD 57189 and 58135 they are opposite. This may be one reason that results in the low DCF peak.

The two X-ray light curves observed by $RXTE$ and $Swift$ have also been analyzed by means of DCF method. The result, DCF(-1.5$^{+1.9}_{-1.9}$) = 0.82$^{+0.04}_{-0.04}$, shows that they are well correlated with each other.

\section{Discussion and Conclusions}

\subsection{Periodicity}

In the 32-yr-long optical $R$ band light curve, we have found there exists a possible 2044-d (5.6-yr) quasi-periodic variation component. The outbursts appear periodically with approximate 5.6-yr intervals (5.79 yr, 4.40 yr, 5.44 yr, 5.45 yr and 5.88 yr).
However, the variation is complex. There are some smaller outbursts in the light curve. The 10-d-averaged light curve has also been adopted to search periodic component. The DCF peak locates on the $\tau$ = 5.5 yr, although the value of DCF drops down to 0.38. This suggest the source varies with a weak quasi-periodicity of $\sim$5.6 yr. Thus it can be predicted that the next outburst maybe happen in 2022 October-November corresponding to the outburst during 2017 March.

On a medium time scale, \cite{li09} detected a 130.6-d outburst periodicity.  However, with ten years of photometric data from 2002 September to 2012 September, \cite{Nilsson18} found no significant periodicity in the optical light curve of 3C 279.
Using optical data between 1929 and 1952, \cite{eachus75} inferred that there exists a tendency to repeat outbursts at $\sim$7 yr intervals.
We have re-examined that light curve. During the 1936-1937, there is a double maximum structure with $B\sim12.10^{m}$ and $B\sim11.27^{m}$ locates on MJD $\sim$28305 and $\sim$28635, respectively, which is separated by an interval of 330 d (0.9 yr). After this, on MJD $\sim$30843, there is another maximum with $B\sim13.29^{m}$. According to two separated bursts, there is an interval of $\sim$6 yr.

Periodic signals have been derived in several sources. For example, an 11-12 yr periodicity for OJ 287 has been detected by \cite{sillanpaa88} and \cite{kidger92b}.
A 317-d periodicity for PKS 2155-304 has been claimed by \cite{zhang14}, and then confirmed by \cite{sandrinelli14}. The evidence of quasi-periods of $\sim$3 and $\sim$1.9 yrs has been found for 3C 66A and B2 1633+38, respectively \citep{otero20}. The phenomenon of quasi-periodicities have been interpreted by several different models. Super-massive binary black hole (SMBBH) model has been proposed to interpret quasi-periodic variations. The secondary black hole maybe change the accretion rate, or lead to the jet precession. Then, the precession cause the variation of the Doppler factor. Helical or helical structure jets also cause the changes in the Doppler factor. There are other mechanisms such as recurrent shock front formation or disk instability, which could produce the quasi-periodical variability in the light curve.  \citep[see][and the references therein]{sandrinelli18, otero20, yang20,agarwal21}.

\subsection{Spectral Index}

The variability in brightness is usually accompanied by that in spectral index. On long time scales, the source 3C 279 shows a complicated optical spectral behavior. In this analysis, the source exhibits considerable variation in optical spectral index and has a mean value of -1.71. \cite{webb90} reported a spectral index variation range of -2.05 $\sim$ -0.55 for 3C279 with an average of -1.12.
Some other reports showed that the optical spectral index of 3C 279 varied from -2.17 to -0.6 \citep{odell78,sitko82,brown89,netzer96,grandi96}.

The optical spectral behavior has been explored for a number of blazars \citep[see][and the references therein]{rani10,zhang15,raiteri17}. In general, there are two typical spectral behaviors. One case is when the brightness increases, the source turns blue (i.e., the spectrum becomes flat). In another case, just the opposite, the source turns red when the brightness increases (i.e., the spectrum becomes steep).
In addition, there are some complex spectral behaviors. For several blazars, when they are in low states, the energy spectra become steeper as the brightness increases. However, when they are in high states, the spectra become flatter or remain unchanged as the brightness increases \citep{villata06,ikejiri11,zhang13,zhang15,isler17}.  For 3C 279, the short-term color variations have been studied, which has been found to show in some cases no apparent correlation between spectral index and brightness \citep{brown89,webb90}, and in other cases the source shows a clear bluer trend when the brightness increases \citep{shrader94,larionov08,rani10,zhang15}.
With over 7 years of SMARTS monitoring, \cite{isler17} found the optical/near-infrared (OIR) color of 3C 279 was averagely bluer when brighter. However, on shorter timescales, the source exhibited different color behaviors, such as bluer when brighter, redder when brighter, and achromatic phenomenon.
In this analysis, although the dots in the diagram of $\alpha - Flux$ are very scattered on the long-time scale, the profile (average trend) exhibits an obvious relation between spectral index and flux (see Fig.~\ref{alpha}). When the $R$ band flux is less than 3 mJy (15.03 mag), 3C 279 becomes blue as brightness increases. If we zoom in the figure, we can see that, in the lowest state (Flux $\leq \sim 0.5$ mJy), the source becomes blue as brightness decreases. These results are in agreement with those presented by \cite{isler17}. However, in the case of the source being brighter than 3 mJy, it remains stable (stable-when-brighter) with a mean value of $\alpha$ $\simeq$ $-1.6$. These phenomenon may be explained by a combination and different contributions of non-thermal jet and thermal accretion disk emission. Redder-when-brighter is due to a faint jet emission and a strong blue disk emission \citep{villata06,ikejiri11,isler17}.  A variable source with constant and bluer color and an underlying redder source in the jet maybe cause bluer-when-brighter phenomenon \citep{ikejiri11}. When the jet is brighter and brighter, the variable component will be much greater the underlying component, and then the color tends to stabilize.

\subsection{Correlation}

\begin{table}
\begin{center}
\caption[]{ Correlation Analysis between Individual Main Outbursts of $\gamma$-ray, X-ray and Optical Bands.}\label{tab:DCF-gxo}


 \begin{tabular}{cccc}
  \hline\noalign{\smallskip}
Outburst &  DCF      & $\tau$ (d) & Bands                    \\
  \hline\noalign{\smallskip}
MJD 58200-58400 &0.66$\pm$0.28 &0   &$\gamma$-ray vs. X-ray\\
MJD 58000-58200 &0.92$\pm$0.27 &0   &$\gamma$-ray vs. X-ray\\
MJD 56900-57300 &0.98$\pm$0.73 &-1  &$\gamma$-ray vs. X-ray\\
MJD 56300-56900 &0.43$\pm$0.15$^{a}$ &0   &$\gamma$-ray vs. X-ray\\
\hline
MJD 58200-58400 &0.91$\pm$0.40 &0   &$\gamma$-ray vs. Optical\\
MJD 58000-58200 &0.92$\pm$0.32 &0   &$\gamma$-ray vs. Optical\\
MJD 57840-58000 &0.79$\pm$0.23 &0   &$\gamma$-ray vs. Optical\\
MJD 57700-57840 &0.89$\pm$0.26 &0   &$\gamma$-ray vs. Optical\\
MJD 56900-57300 &0.50$\pm$0.31$^{b}$ &0   &$\gamma$-ray vs. Optical\\
\hline
MJD 58200-58400 &0.94$\pm$0.55 &0   &X-ray vs. Optical \\
MJD 58000-58200 &0.96$\pm$0.18 &2   &X-ray vs. Optical\\
MJD 57700-58000 &0.96$\pm$0.69 &-2  &X-ray vs. Optical\\
MJD 56900-57300 &0.83$\pm$0.42$^{c}$ &1   &X-ray vs. Optical\\
  \noalign{\smallskip}\hline
\end{tabular}

\footnotesize{$^a$ There is another peak, DCF(10) = 0.46 $\pm$ 0.20.}\\
\footnotesize{$^b$ There is another peak, DCF(7) = 0.59 $\pm$ 0.34.}\\
\footnotesize{$^c$ There is another peak, DCF(7) = 0.97 $\pm$ 0.88.}\\
\end{center}
\end{table}

In this analysis, 3C 279 shows violent variabilities in all three bands of $\gamma$-ray, X-ray and optical $R$ bands. Almost each main outburst in one light curve has corresponding outbursts in the others.

During 1996 January-February, an X-ray outburst ($RXTE$) of 3C 279 showed a good correlation with the $\gamma$-ray flare with no lags \citep{wehrle98}.
However, \cite{hayashida12} indicated that there was no correlation between the variations of X-ray and $\gamma$-ray bands during 2008-2010.
\cite{fraija19} investigated the correlation between $\gamma$-ray and X-ray, and derived a low DCF ($\sim$ 0.05) peak close to zero lag during the flare of 2014 February-April (MJD 56695-56775).  Recently, \cite{larionov20} found the variations of X-ray and $\gamma$-ray bands were well correlated with each other.
On a long time scale, we find the variations are well correlated between $\gamma$-ray and X-ray with no delay in this analysis. Individual outbursts have also been investigated by DCF method, and they show good correlations with each other (see Tab.~\ref{tab:DCF-gxo}).
The simultaneous correlated variability in $\gamma$-ray and X-ray indicates they are approximately co-spatial \citep{wehrle98}. 3C 279 is a low-frequency peaked blazar.
The radiation of radio to UV wavelengths are generated through synchrotron process by relativistic electrons in the jet, and  X-ray and $\gamma$-ray are produced by scattering soft-target photons in or out of the jet.
The X-ray and $\gamma$-ray might be the low and high energy ends of the same inverse-Compton emission component, respectively \citep{hayashida15}.

With the data of 2008-2010, \cite{hayashida12} reported that a possible 10-d delay between the optical and $\gamma$-ray emissions. However, \cite{janiak12} emphasized that a 10-d lag is just a possibility as the result may be misled by poor data quality. \cite{fraija19} also suggested no correlation between  $\gamma$-ray and optical bands during 2008-2010 since the peak of DCF = $\sim$ 0.055 at the lag of 31 d.
\cite{larionov20} suggested that the correlation was rather weak between $\gamma$-ray and optical bands, and the relation changed with activity state.
In this analysis, the variations of $\gamma$-ray and optical $R$ bands show a weak correlation. The low DCF may be due to the disproportion between different outbursts. Then we investigate the individual $\gamma$-ray outbursts respectively. For the outburst during MJD 58200-58400 (2018 March - October), $DCF(0)=0.91 \pm 0.40$.
The DCF results of the individual main outbursts are listed in Table~\ref{tab:DCF-gxo}, which indicate that the variations of $\gamma$-ray have strong correlations with those of optical band, and there are no obvious time delays.
There is evidence to support that optical and $\gamma$-ray emission regions have common spatiality \citep{abdo10}.
In the simplest case, the connection between the flares of optical and gamma-ray bands can be predicted by both synchrotron self-Compton (SSC) and External Compton (EC) models \citep{reinthal12}.
\cite{cohen14} found a clear correlations between optical and $\gamma$-ray variations in 30 out of the 40 brightest blazars, which clearly favored the leptonic model used to explain blazar radiation.

\cite{larionov08} pointed out the X-ray$-$optical correlation of 3C 279 was rather weak during 2006-2007, with variations in X-ray being $\sim$1 d ahead of optical ($R$) variations.
\cite{chatterjee08} also found significant correlations between the variations of X-ray (observed by $RXTE$) and optical $R$ bands during 1996-2007. \cite{fraija19} indicated no correlation between X-ray and optical bands during 2009. In this analysis, for X-ray (observed by $RXTE$) and optical $R$ bands, their variation trends are very similar. They exhibit strong correlation with each other.
However, DCF results have revealed that X-ray (observed by $Swift$) variations are weakly correlated with those of optical band. If we examine the light curves in detail (Fig.~\ref{lc-swift-optical}), we find the source bursts almost simultaneously, and the light curves have similar trends. However, the magnitude of the main flares are not proportionate. We have checked individual bursts of X-ray band. For the maximal outburst of X-ray on MJD 58000-58200 (2017 September to 2018 March), DCF of X-ray and optical bands equals 0.96 at 2 d. The DCF result of other bursts have been investigated and listed in Table~\ref{tab:DCF-gxo}. For each outburst of X-ray and optical bands, the variations are well correlated.
This indicates that the optical flares are emitted by the synchrotron process, the X-ray emission is not simply a extension of synchrotron process, but instead, includes the contribution of inverse Compton scattering components of low energy electrons \citep{larionov08,chatterjee08,abdo10,chatterjee12}.

To conclude, we have compiled the long term optical, X-ray and $\gamma$-ray light curves of blazar 3C 279. The source exhibits violent variabilities in all those three bands. In the optical band, 3C 279 has been found to exist a possible 5.6-yr-long weak quasi-periodic variation component, and exhibit different spectral behaviors in different states, with a clear flatter-when-brighter trend in low states and a stable-when-brighter trend in high states. The correlation analysis concludes that $\gamma$-ray and X-ray variations are strongly correlated with each other, but weakly with those of optical $R$ band. However, in the course of the main outbursts, the variations of $\gamma$-ray, X-ray and optical $R$ bands show strong correlations with each other and with no obvious time lag. The results mean that the optical emission are produced by the synchrotron process, and the X-ray, $\gamma$-ray by inverse Compton scattering, and their emission region are approximately co-spatial.


\begin{acknowledgements}
We thank the support from National Natural Science Foundation of China (NSFC) under
Nos. U1831124 and 11273008, and the Natural Science Foundation of Anhui Province of China 
with Grant No. 1908085MA28. We also thank the $WEBT$ groups for providing their optical data.

\end{acknowledgements}

\label{lastpage}

\end{document}